\documentclass[twocolumn]{aastex63}

\usepackage{textcomp}
\usepackage{hyperref}
\received{September 15, 2019}
\revised{November 2, 2019}
\accepted{November 17, 2019} %\today}
\submitjournal{the Parker Solar Probe \textit{ApJS} Special Issue}

\shorttitle{Solar Wind Modeling at PSP}
\shortauthors{Kim et al.}
\begin{document}

%\linenumbers

\title{PREDICTING THE SOLAR WIND AT PARKER SOLAR PROBE USING AN EMPIRICALLY DRIVEN MHD MODEL}

\correspondingauthor{T. K. Kim}
\email{tae.kim@uah.edu}

\author[0000-0003-0764-9569]{T. K. Kim}
\affiliation{Center for Space Plasma and Aeronomic Research (CSPAR), The University of Alabama in Huntsville, Huntsville, AL 35805, USA}

\author[0000-0002-6409-2392]{N. V. Pogorelov}
\affiliation{Department of Space Science and CSPAR, The University of Alabama in Huntsville, Huntsville, AL 35805, USA}

\author{C. N. Arge}
\affiliation{NASA Goddard Space Flight Center, Greenbelt, MD 20771, USA}

\author{C. J. Henney}
\affiliation{AFRL/Space Vehicles Directorate, Kirtland AFB, Albuquerque, NM 87117, USA}

\author{S. I. Jones-Mecholsky}
\affiliation{NASA Goddard Space Flight Center, Greenbelt, MD 20771, USA}
\affiliation{Catholic University of America, Washington, DC 20064, USA}

\author{W. P. Smith}
\affil{Middle Tennessee State University, Murfreesboro, TN 37132, USA}

\author[0000-0002-1989-3596]{S. D. Bale}
\affil{Physics Department and Space Sciences Laboratory, University of California, Berkeley, CA 94720, USA}
\affil{The Blackett Laboratory, Imperial College London, London, SW7 2AZ, UK}
\affil{School of Physics and Astronomy, Queen Mary University of London, London E1 4NS, UK}

\author[0000-0002-0675-7907]{J. W. Bonnell}
\affil{Space Sciences Laboratory, University of California, Berkeley, CA 94720, USA}

\author[0000-0002-4401-0943]{T. {Dudok de Wit}}
\affil{LPC2E, CNRS and University of Orl\'eans, Orl\'eans, France}

\author[0000-0003-0420-3633]{K. Goetz}
\affil{School of Physics and Astronomy, University of Minnesota, Minneapolis, MN 55455, USA}

\author[0000-0002-6938-0166]{P. R. Harvey}
\affil{Space Sciences Laboratory, University of California, Berkeley, CA 94720-7450, USA}

\author[0000-0003-3112-4201]{R. J. MacDowall}
\affil{Solar System Exploration Division, NASA/Goddard Space Flight Center, Greenbelt, MD, 20771}

\author[0000-0003-1191-1558]{D. M. Malaspina}
\affil{Laboratory for Atmospheric and Space Physics, University of Colorado, Boulder, CO 80303, USA}

\author[0000-0002-1573-7457]{M. Pulupa}
\affil{Space Sciences Laboratory, University of California, Berkeley, CA 94720-7450, USA}

\author[0000-0002-7077-930X]{J. C. Kasper}
\affiliation{Climate and Space Sciences and Engineering, University of Michigan, Ann Arbor, MI 48109, USA}
\affiliation{Smithsonian Astrophysical Observatory, Cambridge, MA 02138, USA}

\author[0000-0001-6095-2490]{K. E. Korreck}
\affiliation{Smithsonian Astrophysical Observatory, Cambridge, MA 02138, USA}

\author[0000-0002-7728-0085]{M. Stevens}
\affiliation{Smithsonian Astrophysical Observatory, Cambridge, MA 02138, USA}

\author[0000-0002-3520-4041]{A. W. Case}
\affiliation{Smithsonian Astrophysical Observatory, Cambridge, MA 02138, USA}

\author[0000-0002-7287-5098]{P. Whittlesey}
\affiliation{University of California, Berkeley, CA 94720, USA}

\author{R. Livi}
\affiliation{University of California, Berkeley, CA 94720, USA}

\author{D. E. Larson}
\affiliation{University of California, Berkeley, CA 94720, USA}

\author[0000-0001-6038-1923]{K. G. Klein}
\affiliation{School of Physics and Astronomy, University of Arizona, Tucson, AZ 85721, USA}

\author[0000-0002-4642-6192]{G. P. Zank}
\affiliation{Department of Space Science and CSPAR, The University of Alabama in Huntsville, Huntsville, AL 35805, USA}



\begin{abstract}

%\begin{linenumbers}
Since the launch on 2018/08/12, Parker Solar Probe (PSP) has completed its first and second orbits around the Sun, having reached down to 35.7 solar radii at each perihelion. In anticipation of the exciting new data at such unprecedented distances, we have simulated the global 3D heliosphere using an MHD model coupled with a semi-empirical coronal model using the best available photospheric magnetograms as input. We compare our heliospheric MHD simulation results with \textit{in situ} measurements along the PSP trajectory from its launch to the completion of the second orbit, with particular emphasis on the solar wind structure around the first two solar encounters. Furthermore, we show our model prediction for the third perihelion, which occurred on 2019/09/01. Comparison of the MHD results with PSP observations provides a new insight on the solar wind acceleration. Moreover, PSP observations reveal how accurately the ADAPT-WSA predictions work throughout the inner heliosphere.
%\end{linenumbers}

\end{abstract}

\keywords{Sun: heliosphere --- Sun: magnetic fields --- solar wind --- methods: numerical --- magnetohydrodynamics (MHD)}


\section{Introduction} \label{sec:intro}

Launched at 2018/08/12 07:31 UT, Parker Solar Probe (PSP) has become the first spacecraft to probe the solar wind below 0.3 astronomical units (au) on its approach to the first perihelion at 35.7 solar radii (R${_s}$) on 2018/11/06 03:27 UT \citep{Fox2016SSRv}. Using gravity assists from 7 Venus flybys, the spacecraft is projected to reach below 10 R$_{s}$ during the 22nd orbit in late 2024. PSP has already completed its first two orbits with all instruments fully operational as we anticipate the public release of a wealth of exciting new data from near the Sun.

The main science objectives of the PSP mission are to improve the understanding of the heating and acceleration of the solar corona and wind, verify the structure and dynamics of the plasma and magnetic field near the Sun, and determine how energetic particles are accelerated and transported \citep{Fox2016SSRv}. To enable its investigation, PSP is equipped with a suite of instruments, namely the Fields Experiment (FIELDS), Integrated Science Investigation of the Sun (IS$\odot$IS), Wide-field Imager for Solar Probe (WISPR), and Solar Wind Electrons Alphas and Protons (SWEAP). FIELDS measures the electric and magnetic fields and waves, Poynting flux, absolute plasma density and electron temperature, spacecraft floating potential and density fluctuations, and radio emissions \citep{Bale2016SSRv}. IS$\odot$IS observes energetic electrons, protons and heavy ions that are accelerated to high energies (10s of keV to 100 MeV) in the sun's atmosphere and inner heliosphere \citep{McComas2016SSRv}. WISPR takes coronagraph-like images of the solar corona and inner heliosphere, and also images of the solar wind, shocks and other structures as they approach and pass the spacecraft, which complement the direct measurements from other instruments by imaging the plasma they sample \citep{Vourlidas2016SSRv}. SWEAP counts the electrons, protons and helium ions and determines the bulk properties such as velocity, density, and temperature \citep{Kasper2016SSRv}.

Three-dimensional (3D), time-dependent solar wind models can be an invaluable tool to support and add context to the single-point observations of interplanetary magnetic field and plasma made by the FIELDS and SWEAP instruments along the highly elliptical PSP orbit. Making predictions for periods of particular interest, such as Venus flybys and perihelia, has become a popular topic in the heliophysics modeling community since launch. For example, \cite{vanderHolst2019ApJL} used the Alfven Wave Solar atmosphere Model (AWSoM) to predict that PSP would cross the heliospheric current sheet two times while sampling mostly slow wind streams (360-420 km s$^{-1}$) during a 12-day period centered around the first perihelion. On the other hand, \cite{Riley2019ApJ} used the Magnetohydrodynamic Algorithm outside a Sphere (MAS) code with a different empirical input to predict only one current sheet crossing during the same period as AWSoM. It is interesting to note that the MAS predictions of the solar wind speed, density, and radial magnetic field strengths also largely disagree with the AWSoM predictions \citep{Riley2019ApJ}. Clearly, solar wind models can differ greatly depending on the numerical approach and source of boundary conditions they employ.

In the following section, we describe our own MHD solar wind model and empirically derived boundary conditions used in this study. Then we present the model results compared with hourly-averaged PSP FIELDS and SWEAP data for the first and second orbits, as well as providing a prediction for the third perihelion.

\begin{figure}[ht]
\begin{center}
\noindent\includegraphics[width=1.0\columnwidth, angle=0]{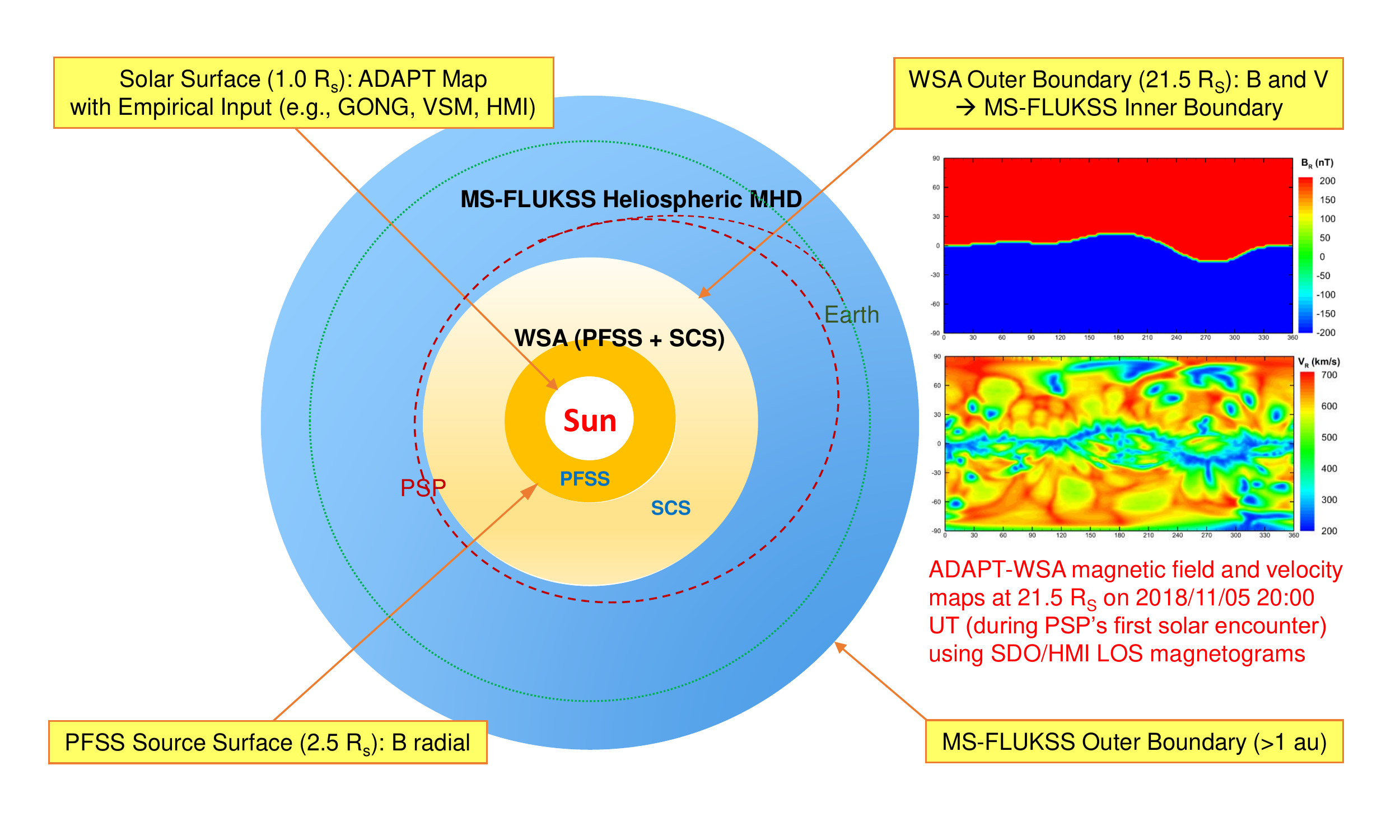}
\end{center}
\caption{Diagram showing the time-dependent model used in this study. WSA model consisting of PFSS and SCS components use ADAPT maps at the solar surface as input to provide radial magnetic field and velocity at the inner boundary of MS-FLUKSS. The trajectories of PSP and Earth are also shown (not to scale).}
\label{fig1}
\end{figure}

\section{MS-FLUKSS Model with HMI-ADAPT-WSA Maps} \label{sec:highlight}

To simulate the 3D, time-dependent variations in the solar wind along the trajectory of PSP, we use the Multi-scale Fluid-kinetic Simulation Suite (MS-FLUKSS), which is a package of numerical codes designed to model the flows of partially ionized plasma in multiple scales with high resolution on a Cartesian or spherical grid using adaptive mesh refinement \cite[see][and references therein]{Pogorelov2014XSEDE}. As illustrated in Figure \ref{fig1}, the MS-FLUKSS heliospheric MHD model is coupled with the Wang-Sheeley-Arge (WSA) coronal model at the heliocentric distance of 21.5 $R_{s}$ (0.1 au). The WSA model is a semi-empirical coronal model for the ambient solar wind \citep{Arge2003AIP,Arge2004JASTP,Arge2005ASP} consisting of a magnetostatic potential field source surface (PFSS) \citep{AltschulerNewkirk1969SoPh,Schatten1969SoPh,WangSheeley1992ApJ} and the Schatten current sheet (SCS) \citep{Schatten1971CE} components, which extrapolate the solar magnetic field from the photosphere to a source surface (typically placed at 2.5 $R_{s}$) and then to larger distances while preserving the large-scale current sheet structure. For this study, we set the WSA outer boundary at 21.5 $R_{s}$, where the solar wind speed is estimated using an empirical formula \cite[e.g.,][]{Arge2003AIP,Arge2004JASTP} based on the flux tube expansion factor $f_{s}$ and the minimum angular distance $d$ between the open field footpoint and the nearest coronal hole boundary at the photosphere. The WSA solar wind speed at 21.5 $R_{s}$ are prescribed as follows:
\begin{equation}
V = 285.0 + 625.0/(1.0 + f_{s})^{\alpha} (\beta - \gamma e^{-(d/w)^{\delta}})^{3.0},
\label{eq1}
\end{equation}
where $\alpha = 1/4.5$, $\beta = 1.0$, $w = 2.0$, $\gamma = 0.8$, $\delta = 2.0$ and
\begin{equation}
f_{s} = (R_{ph}/R_{ss})^{2}(B_{ph}/B_{ss}),
\label{eq2}
\end{equation}
where $R_{ph} = 1 R_{s}$, $R_{ss} = 2.5 R_{s}$, and $B_{ph}$ and $B_{ss}$ are the magnetic field strengths at the photosphere and the source surface along each flux tube, respectively. The coefficients in Eq.(\ref{eq1}) have been optimized for the particular empirical input to the WSA model we used in this study, which is described next. With the exception of fixed $\beta$ and $\gamma$, the optimal coefficients can vary for different sources of model input \cite[e.g.,][]{Riley2015SW}. While some recent studies suggest that a lower source surface height may be more realistic for solar cycle 24 \cite[e.g.,][]{Nikolic2019SW,Szabo2019ApJS}, we maintain the PFSS source surface (i.e., $R_{ss}$ in Eq.(\ref{eq2})) at the traditional height of 2.5 $R_{s}$.

\begin{figure*}[ht]
\begin{center}
%\begin{array}{cc}
\noindent\includegraphics[width=0.49\textwidth, angle=0]{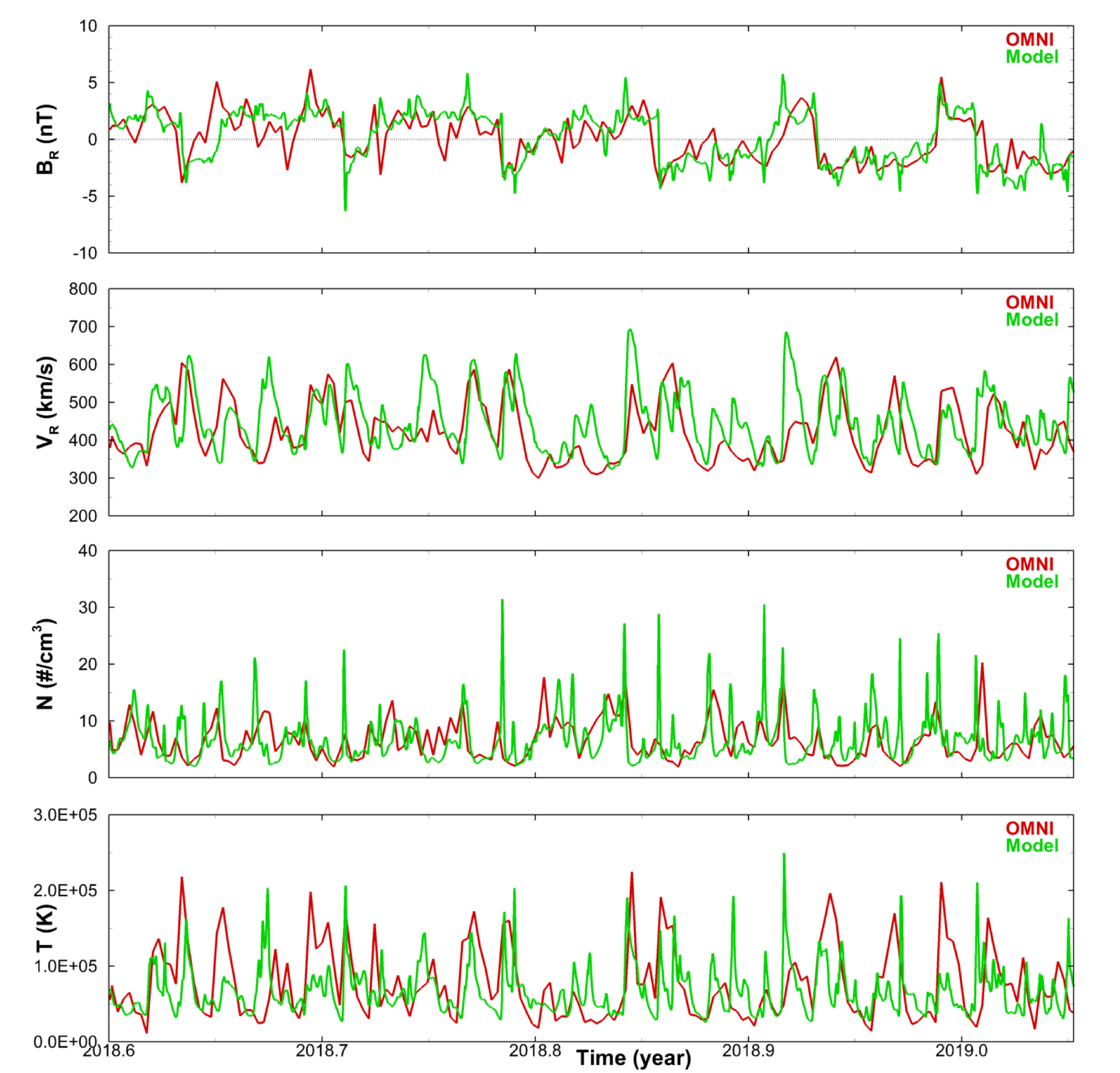}
\noindent\includegraphics[width=0.49\textwidth, angle=0]{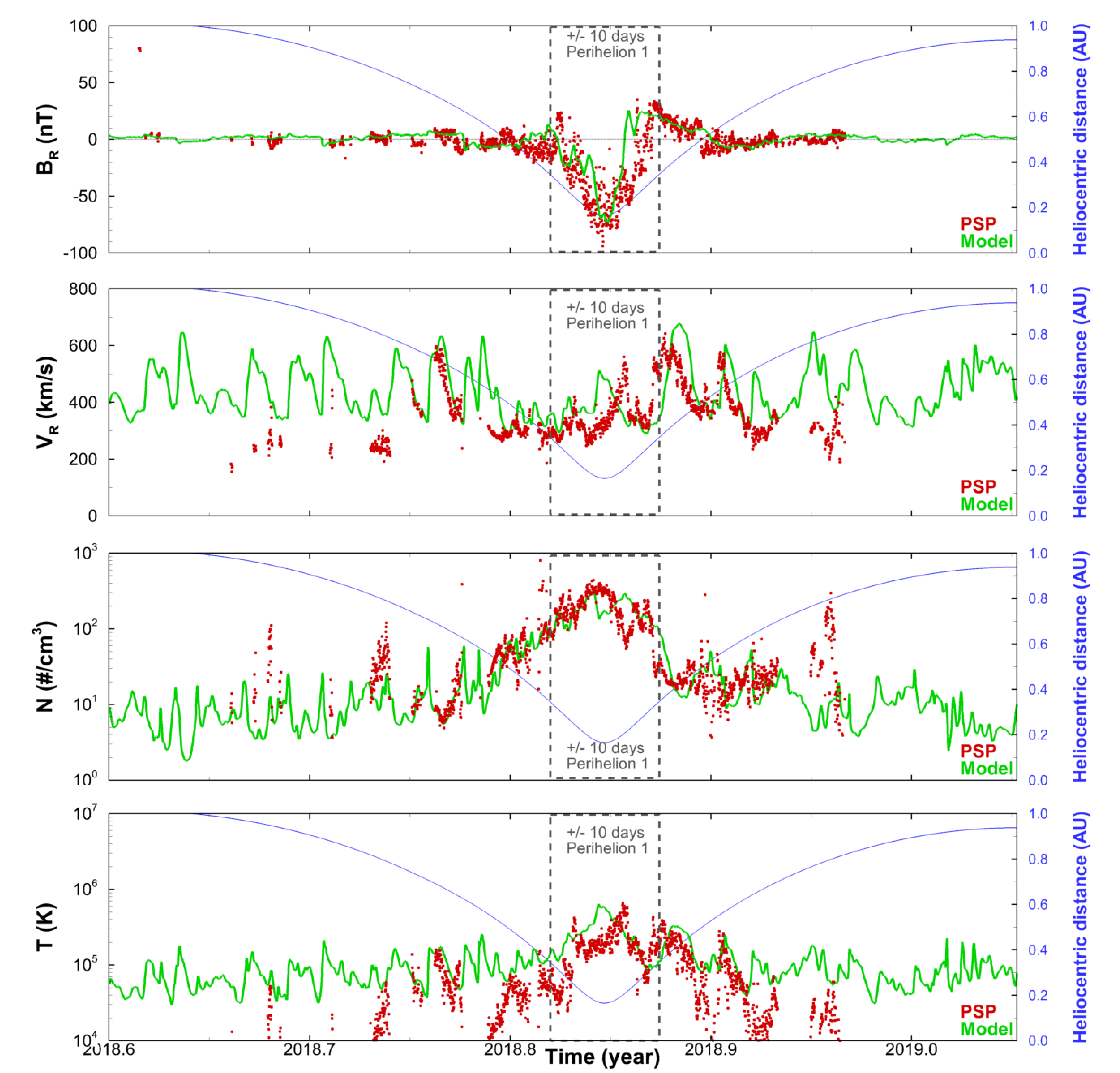}
%\end{array}$
\end{center}
\caption{Radial components of magnetic field (nT) and solar wind velocity (km s$^{-1}$), proton density (cm$^{-3}$) and temperature (K) at Earth (left column) and PSP (right column) during the first orbit of PSP. Model results are shown in green while near-Earth (OMNI) and PSP FIELDS and SWEAP data are shown in red.}
\label{fig2}
\end{figure*}

The WSA model considers various sources of input at the solar surface, such as synoptic NSO/GONG magnetograms and the Air Force Data Assimilative Photospheric flux Transport (ADAPT) model that provides a time sequence of synchronic maps by assimilating NSO/SOLIS/VSM, GONG or SDO/HMI line-of-sight magnetograms into a flux-transport model using localized ensemble Kalman filtering techniques \citep{Arge2010AIP,Arge2011ASP,Arge2013AIP,Hickmann2015SoPh}. In the case of VSM magnetograms, for example, \cite{Hickmann2015SoPh} estimate the observational error to be 3\% that increases sharply towards the limb, where more weight is applied to the ADAPT model values during data assimilation. We note that magnetograph observations from different instruments can vary by up to a factor of 2 \citep{Riley2007ApJ}. To drive the MS-FLUKSS heliospheric MHD model, we select one particular realization (out of 12) of HMI-ADAPT-WSA output for each PSP orbit that provides the best agreement with near-Earth solar wind data compared to synoptic GONG-WSA results or other ADAPT-WSA realizations employing different sources of input magnetograms. We currently rely on visual inspection to qualitatively determine the best sequence of WSA maps, but we may be able to use a newly-developed, quantitative ranking procedure in future studies.

While the WSA model assumes magnetic field to be entirely radial at its outer boundary, an azimuthal component develops in the inertial coordinate system of MS-FLUKSS due to Sun's rotation. Hence, we estimate the azimuthal component using the local solar wind speed to allow for the Sun's rotation and adjust the radial component to conserve the original WSA magnetic flux \cite[e.g.,][]{MacNeice2011SW}. Before interpolating the original 2\textdegree $\times$ 2\textdegree\ ($\phi$, $\theta$) WSA maps onto the MS-FLUKSS inner boundary, we scale the WSA magnetic field uniformly by a factor of 2 to compensate for the systematic underestimation of the magnetic field strengths at 1 au \cite[e.g.,][]{Linker2016JPCS,Linker2017ApJ,Wallace2019SoPh}. The WSA solar wind speeds are also reduced by 75 km s$^{-1}$ to account for the differences in solar wind acceleration between the simple kinematic model of WSA and the more sophisticated MS-FLUKSS MHD model \cite[e.g.,][]{MacNeice2011SW,Kim2014JGR}. We further estimate the solar wind density and temperature at the MS-FLUKSS inner boundary based on the assumptions of constant momentum flux and thermal pressure balance, respectively \cite[see][and references therein]{Linker2016JPCS}. An example of the radial magnetic field and solar wind velocity at the WSA/MS-FLUKSS interface is shown in Figure \ref{fig1}.

Using the time sequence of magnetic field and solar wind velocity, density, and temperature for 2018/08/01 20:00:00 UT - 2019/08/13 20:00:00 UT derived from the WSA model as inner boundary conditions, we solve the ideal MHD equations on a nonuniform $200 \times 256 \times 128$ ($r$, $\phi$, $\theta$) spherical grid (e.g., $\Delta$r $\approx$ 0.645, 1.08, and 1.72 $R_{s}$ at r = 0.1, 0.5 and 1.0 au, respectively), with the specific heat ratio $\gamma$ set to 1.5. While MS-FLUKKS allows the user to model the interaction between the solar wind and the local interstellar medium out to hundreds or even many thousands of au from the Sun \cite[e.g.,][]{Pogorelov2015ApJL}, we set the outer boundary at 1.1 au to focus on the trajectory of PSP that lies entirely within the inner heliosphere.

\section{Results} \label{sec:cite}

\begin{figure*}[ht]
\begin{center}
%\begin{array}{cc}
\noindent\includegraphics[width=0.58\textwidth, angle=0]{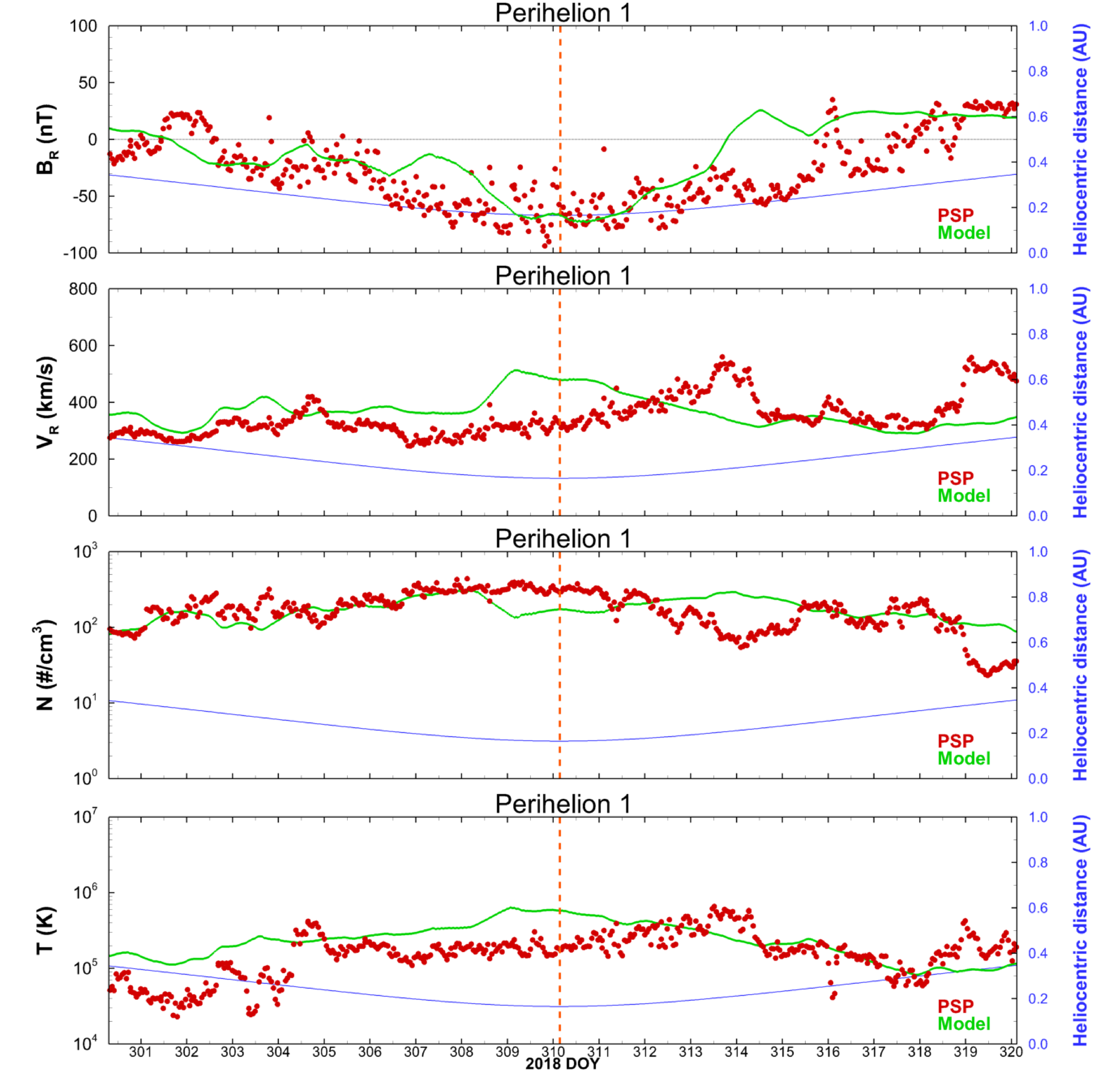}
\noindent\includegraphics[width=0.41\textwidth, angle=0]{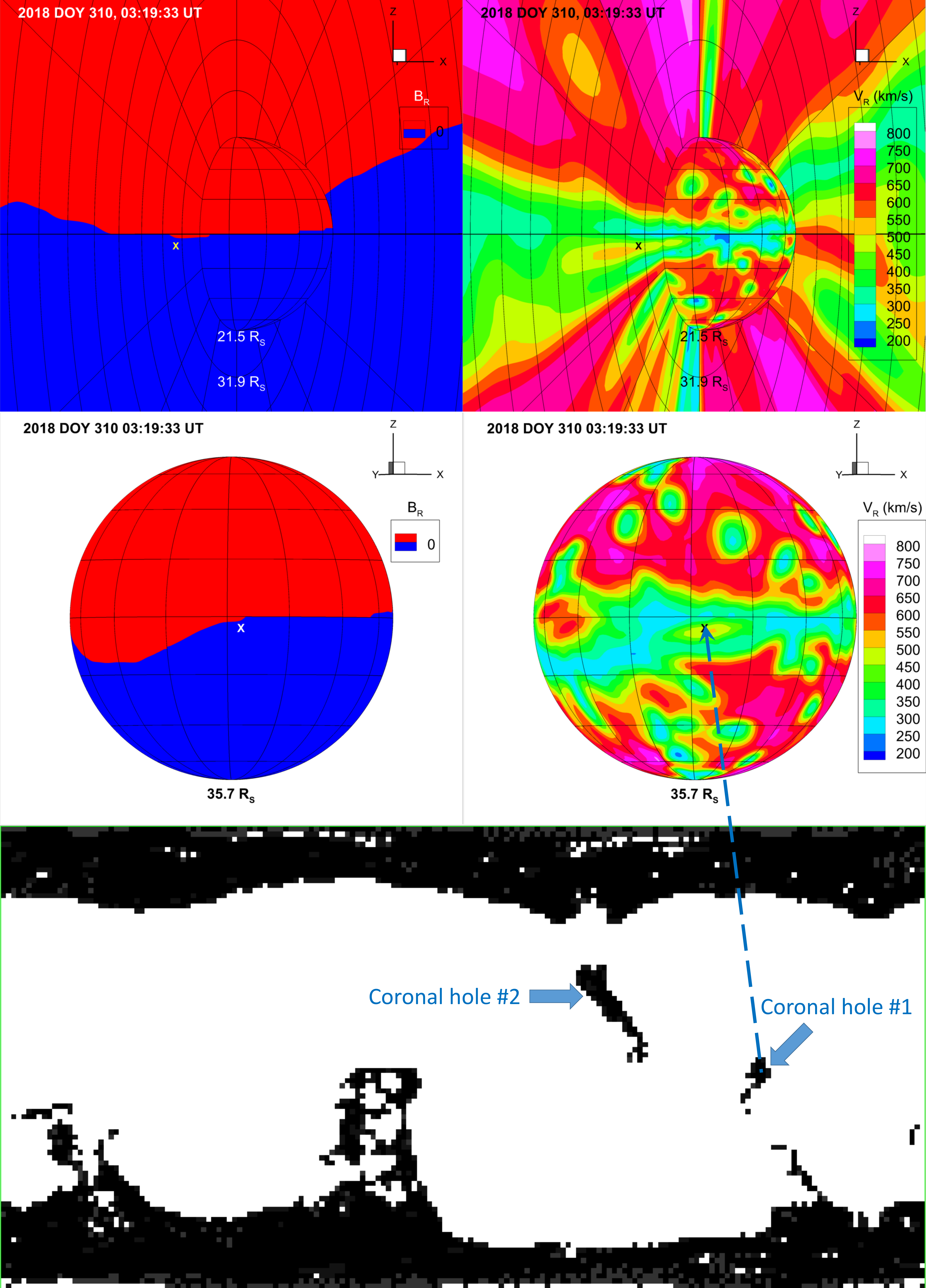}
%\end{array}$
\end{center}
\caption{Left panel: Radial components of magnetic field (nT) and solar wind velocity (km s$^{-1}$), proton density (cm$^{-3}$) and temperature (K) at PSP within +/-10 days of the first perihelion, which is marked by a vertical dashed line. Right panel: Radial components of magnetic field and solar wind velocity shown in 3D (top row) and on a spherical slice at the perihelion distance of 35.7 R$_{s}$ (middle row) on 2018/11/06 (DOY 310) 03:19:33 UT, where a dashed line connects the PSP location marked by an X to the source region in the coronal hole map on the photosphere (bottom).}
\label{fig3}
\end{figure*}

\begin{figure*}[ht]
\begin{center}
%\begin{array}{cc}
\noindent\includegraphics[width=0.49\textwidth, angle=0]{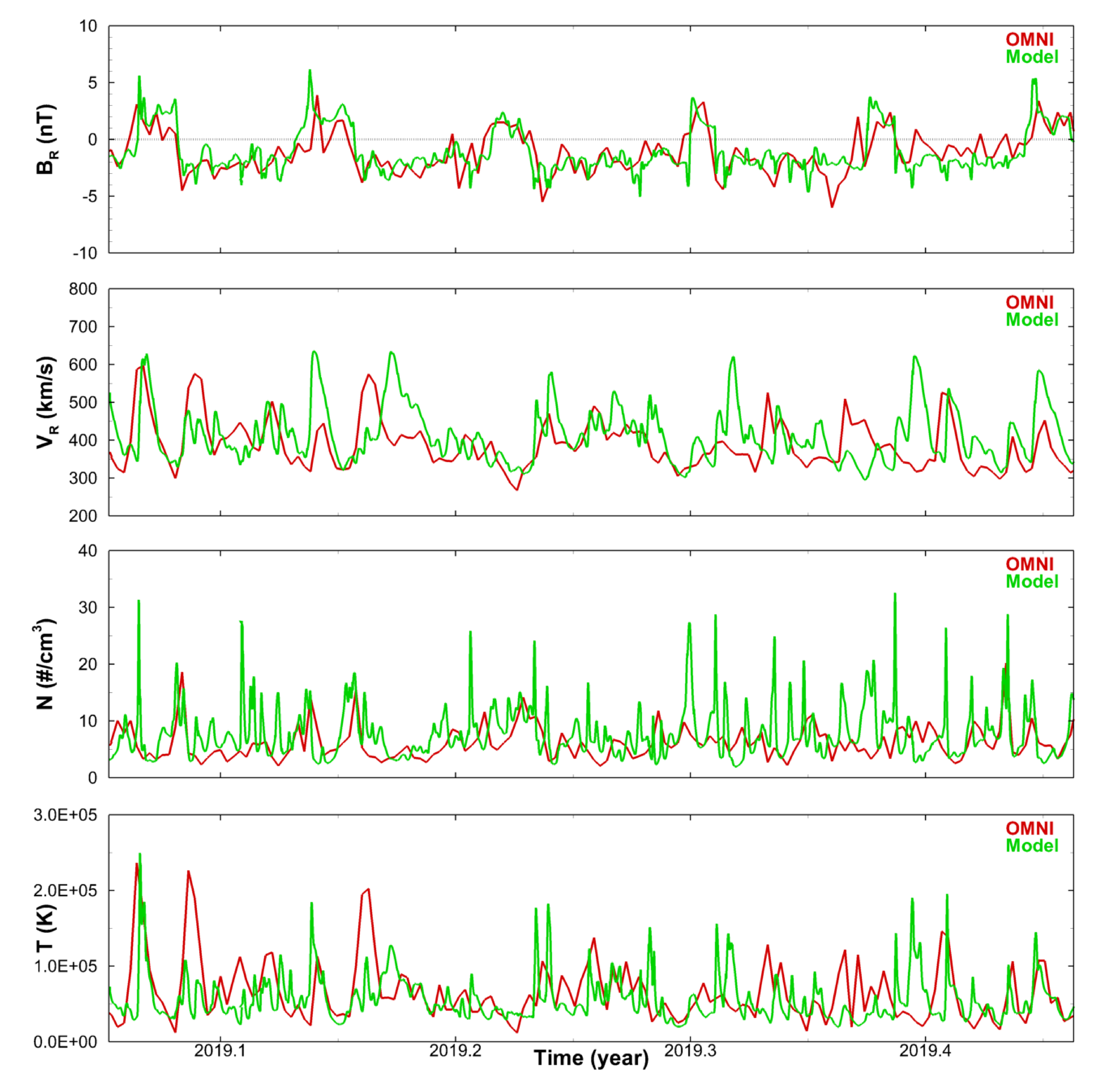}
\noindent\includegraphics[width=0.49\textwidth, angle=0]{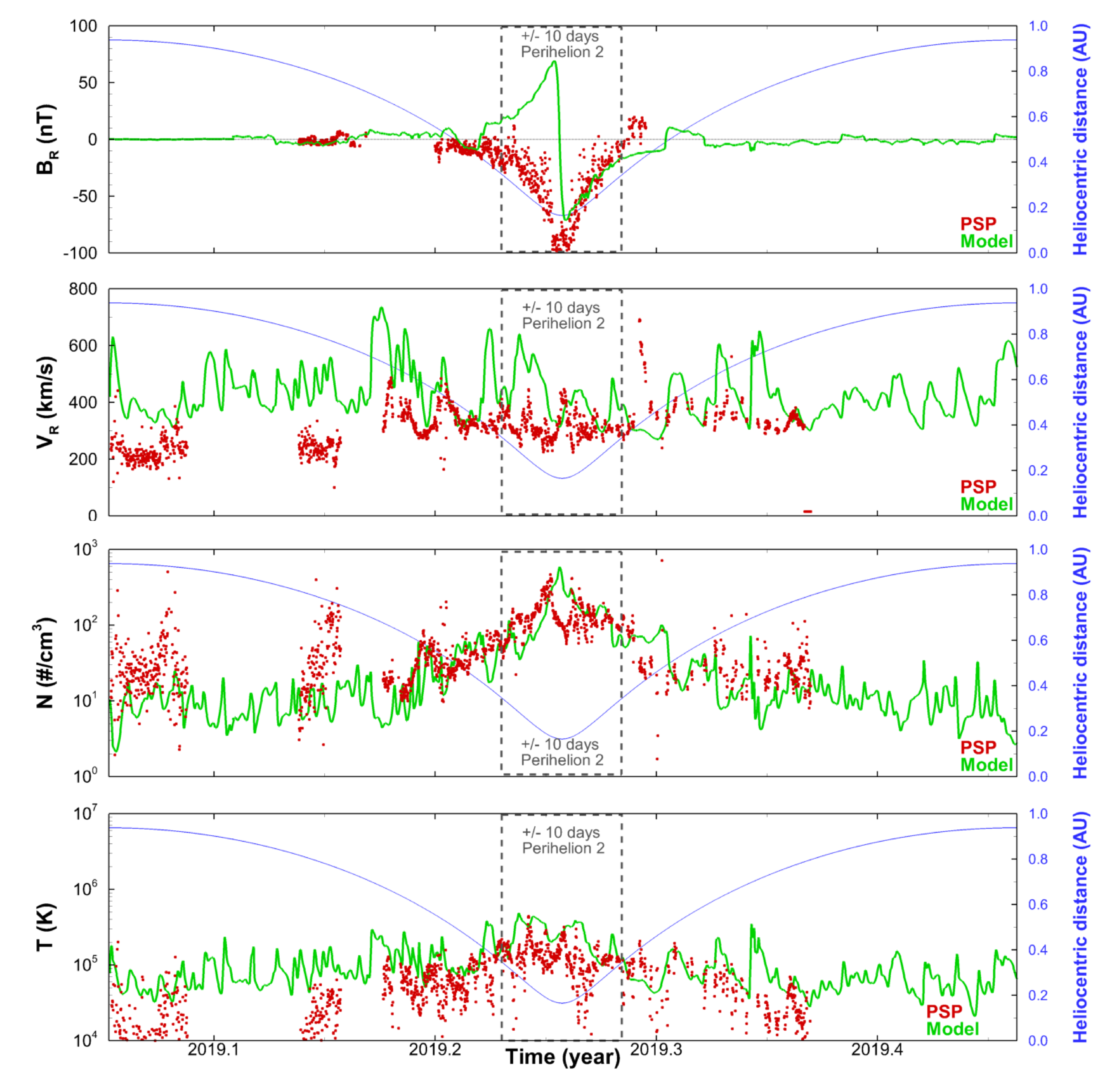}
%\end{array}$
\end{center}
\caption{Radial components of magnetic field (nT) and solar wind velocity (km s$^{-1}$), proton density (cm$^{-3}$) and temperature (K) at Earth (left panel) and PSP (right panel) during the second orbit of PSP. Model results are shown in green while near-Earth (OMNI) and PSP FIELDS and SWEAP data are shown in red. The PSP comparisons also show the heliocentric distance in blue.}
\label{fig4}
\end{figure*}

\begin{figure*}[ht]
\begin{center}
%\begin{array}{cc}
\noindent\includegraphics[width=0.58\textwidth, angle=0]{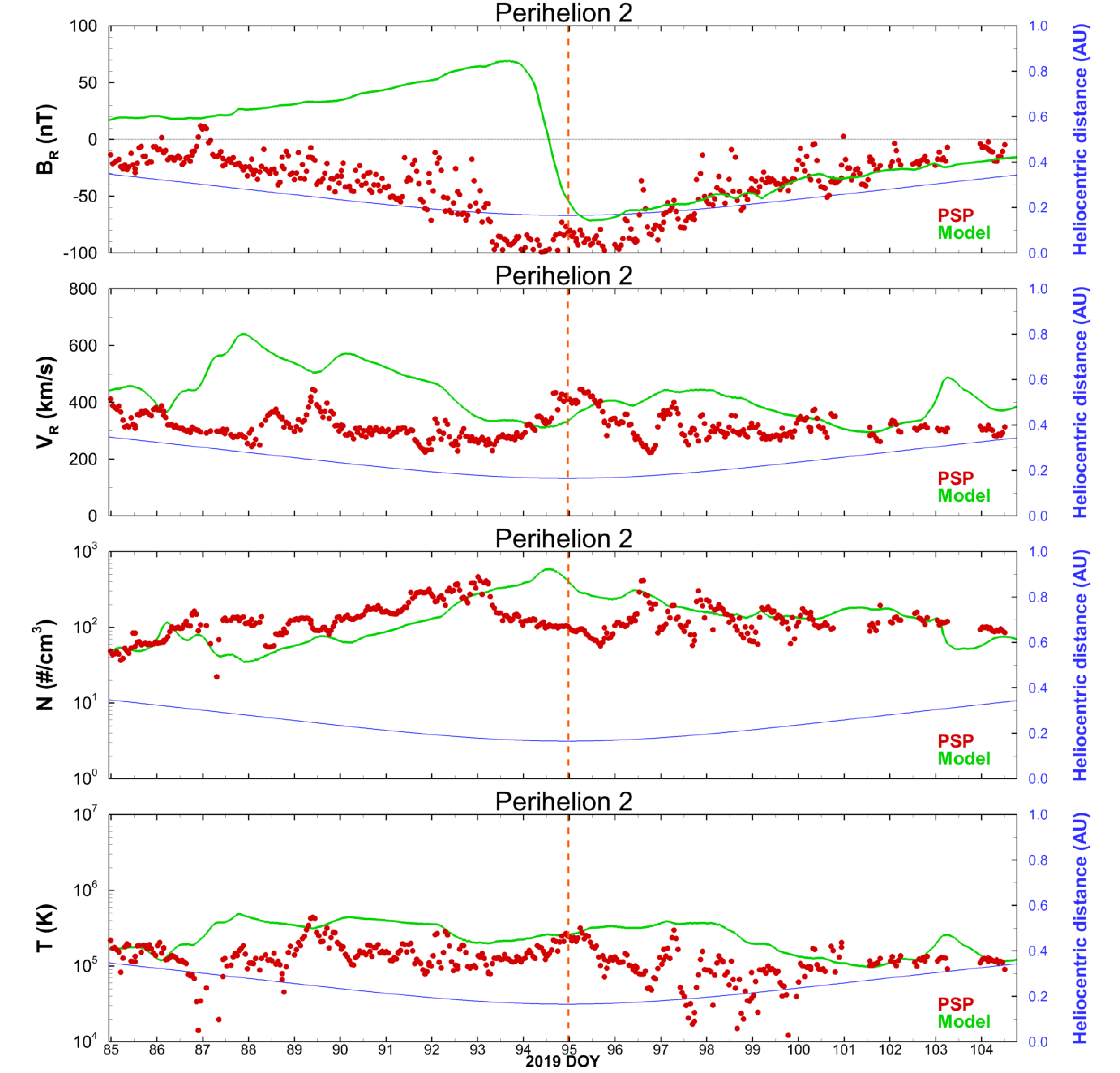}
\noindent\includegraphics[width=0.41\textwidth, angle=0]{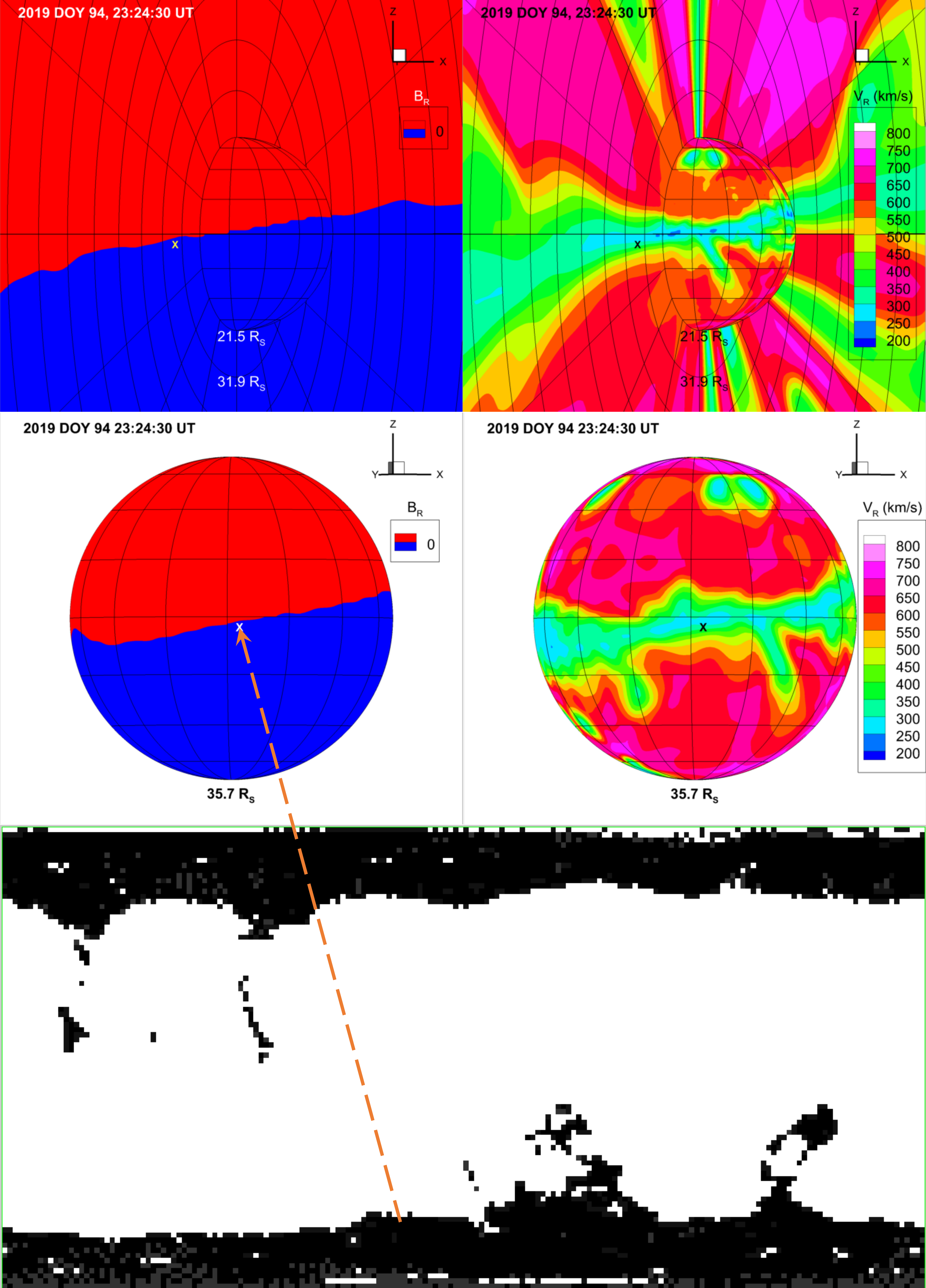}
%\end{array}$
\end{center}
\caption{Left panel: Radial components of magnetic field (nT) and solar wind velocity (km s$^{-1}$), proton density (cm$^{-3}$) and temperature (K) at PSP within +/-10 days of the second perihelion, which is marked by a dashed line. Right panel: Radial components of magnetic field and solar wind velocity shown in 3D (top row) and on a spherical slice at the perihelion distance of 35.7 R$_{s}$ (middle row) on 2019/04/04 (DOY 94) 23:24:30 UT, where a dashed line connects the PSP location marked by an X to the source region in the coronal hole map on the photosphere (bottom).}
\label{fig5}
\end{figure*}

\begin{figure}[ht]
\begin{center}
\noindent\includegraphics[width=1.0\columnwidth, angle=0]{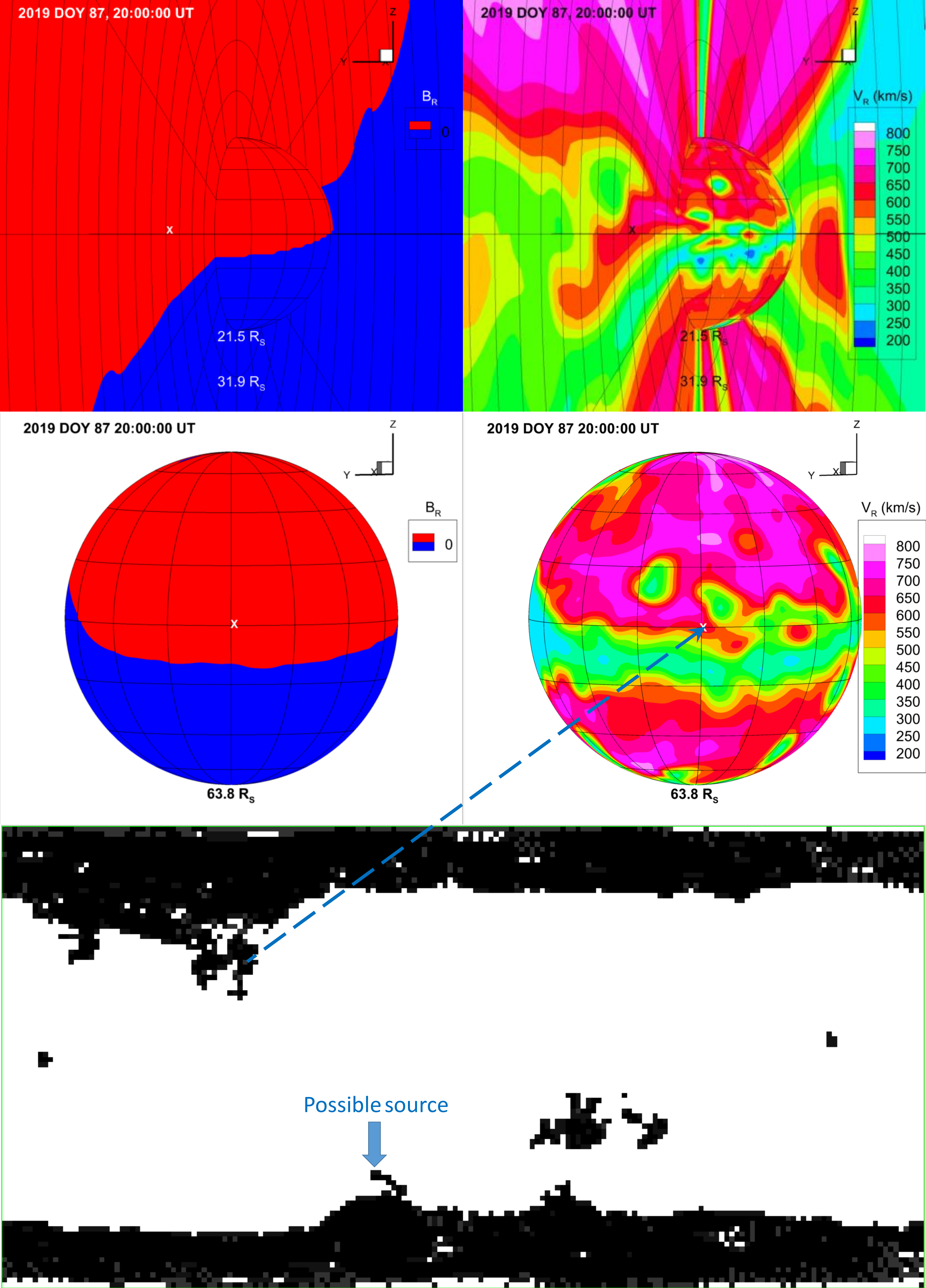}
\end{center}
\caption{Radial components of magnetic field and solar wind velocity shown in 3D (top row) and on a spherical slice at the PSP distance of 63.8 R$_{s}$ (middle row) on 2019/03/28 (DOY 87) 20:00:00 UT, where a dashed line connects the PSP location marked by an X to the source region in the coronal hole map on the photosphere (bottom).}
\label{fig6}
\end{figure}

\begin{figure*}[ht]
\begin{center}
%\begin{array}{cc}
\noindent\includegraphics[width=0.58\textwidth, angle=0]{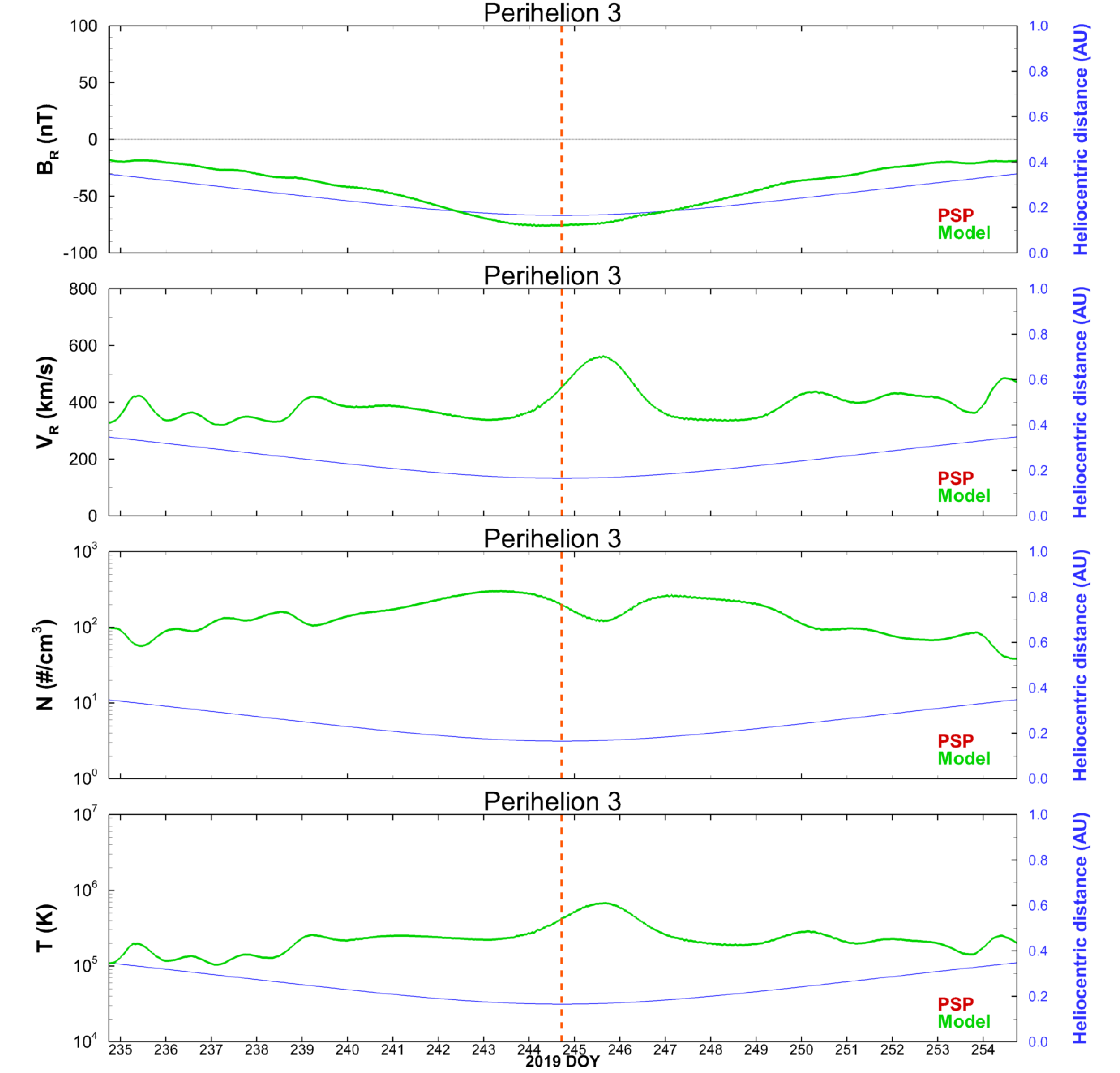}
\noindent\includegraphics[width=0.41\textwidth, angle=0]{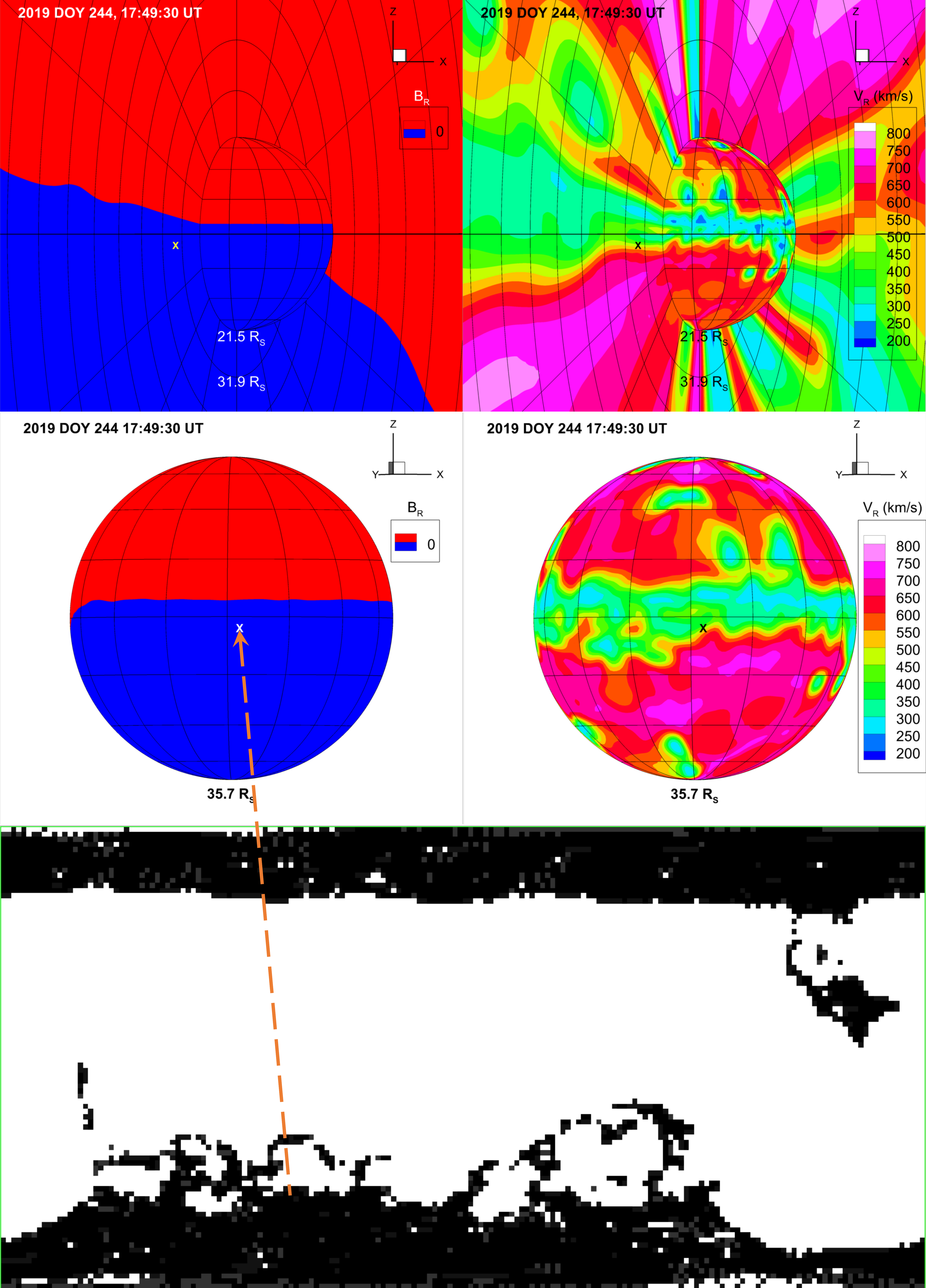}
%\end{array}$
\end{center}
\caption{Left panel: Radial components of magnetic field (nT) and solar wind velocity (km s$^{-1}$), proton density (cm$^{-3}$) and temperature (K) at PSP within +/-10 days of the third perihelion, which is marked by a dashed line. Right panel: Radial components of magnetic field and solar wind velocity shown in 3D (top row) and on a spherical slice at the perihelion distance of 35.7 R$_{s}$ (middle row) on 2019/09/01 (DOY 244) 17:49:30 UT, where a dashed line connects the PSP location marked by an X to the source region in the coronal hole map on the photosphere (bottom). This simulation was performed on 2019/08/31 using the last available HMI-ADAPT-WSA map from 2019/08/13 20:00 UT. The PSP data for Orbit 3 will be made ‘public’ after Orbit 4 data is fully downlinked sometime in 2020.}
\label{fig7}
\end{figure*}

Figure \ref{fig2} shows the radial components of the model magnetic field and velocity, proton density and temperature compared with OMNI data \citep{King2005JGR} at Earth and PSP data for the first orbit around the Sun from 2018/08/12 to 2019/01/19. Between 2018.60 and 2018.65, the model compares reasonably to OMNI data, which suggest a fast wind stream of negative magnetic polarity preceded by a slow wind stream of positive magnetic polarity. However, there is a considerable discrepancy around 2018.65 when a coronal mass ejection (CME) arrived at Earth and caused a strong geomagnetic storm. Despite the predominantly quiet solar wind conditions as the solar minimum approaches, there have been a few CMEs in Earth's direction since the launch of PSP. The WSA model only provides information about the large-scale ambient solar wind, so it is not realistic to expect our model to agree with OMNI data during CME passages. Though it is possible to simulate each individual CME in the ambient solar wind that our model generates \cite[e.g.,][]{Singh2018ApJ,Singh2019ApJ}, we disregard CMEs in this study to focus on the general, large-scale variations in the solar wind along the PSP trajectory.

Between 2018.65 and 2019.05, the model reproduces the overall sector structure at Earth reasonably except around 2018.67, 2018.68, 2018.73, and 2018.76, where the model suggests magnetic field of positive polarity in contrast to the transient flip to negative polarity in OMNI data. Those times are marked by the presence of a number of Earth-facing equatorial coronal holes and northward extension of the southern polar coronal hole, and it is possible that some of these features may not have been reproduced accurately by the WSA model. The model radial velocity also generally agrees with the fluctuations in OMNI data, with notable deviations around 2018.67, 2018.75, 2018.85, and 2018.92, where the model overestimates the variations by at least 150 km s$^{-1}$. The comparisons of proton density and temperature exhibit similar trends because the accuracy of those parameters largely depend on that of the solar wind speed.

In the first half of the period up to 2018.79, the model radial magnetic field and velocity at PSP fluctuate mostly in the -10 to +10 nT and 400 to 600 km s$^{-1}$ ranges, respectively, as the heliocentric distance gradually decreases to 0.5 au. During the first solar encounter, the model radial magnetic field decreases to -80 nT, which agrees remarkably with the observed amplitudes, while velocity fluctuation reduces to the 300 to 400 km s$^{-1}$ range until 2018.84. After the first solar encounter, the model radial field and velocity again fluctuate mostly in the -10 to +10 nT and 400 to 600 km s$^{-1}$ range as PSP gradually approaches the first aphelion. The model proton density and temperature steadily increase from 2-20 cm$^{-3}$ and $5 \times 10^{4}-2 \times 10^{5}$ K near 1 au to 100-300 cm$^{-3}$ and 1-6$\times 10^{5}$ K during the first solar encounter. These results are mostly consistent with the PSP FIELDS and SWEAP data excluding comparison at distances much larger than 0.25 au since the SWEAP measurements are frequently made at low signal/noise beyond that distance and thus may contain artifacts.

The left panel of Figure \ref{fig3} provides an expanded view of the simulation results at PSP for the first solar encounter during the 20-day period around perihelion 1 (2018/11/06 03:27 UT). The model suggests that PSP crosses the heliospheric current sheet from positive to negative magnetic polarity at 2018/10/28 15:40 UT (DOY 301.6) and back to positive at 2018/11/09 18:30 UT (DOY 313.8). Similarly, the FIELDS data also show that PSP encountered mostly negative magnetic polarity for at least two weeks centered around the perihelion, except during a CME passage on 2018/11/11-2018/11/12 (DOY 315-316) when the magnetic polarity briefly switches to positive. Both the model and PSP data indicate that the radial velocity fluctuates mainly between 300 and 400 km s$^{-1}$, except for two high-speed streams above 500 km s$^{-1}$ at DOY 309 and 322 in the model and DOY 313 and 319 in the SWEAP data. The high-speed stream at DOY 309 persists over the perihelion at DOY 310 in the model while a similar high-speed stream is observed by PSP three days after the perihelion. The right panel of Figure \ref{fig3} shows a snapshot of 3D and spherical slices (35.7 R$_{s}$ or 0.166 au) of the model magnetic field polarity and radial velocity at the perihelion, where the location of PSP is marked by an `x'. These plots suggest that PSP was within 2\textdegree\ of the heliospheric current sheet, which is traced by the boundary between the red (positive)) and blue (negative) colors in the magnetic polarity plots. The radial velocity plots show that PSP was traversing the edge of a high speed stream of negative magnetic polarity connected to an equatorial coronal hole in the southern hemisphere, which is labeled as coronal hole \#1 in the photosphere map at the bottom. It appears that, in the model, PSP may have grazed this stream 3-4 days too early and thus crossed the heliospheric current sheet 3-4 days prematurely as well. On the other hand, the second high-speed stream at DOY 319 in the SWEAP data, which is of positive magnetic polarity, appear at DOY 322 in the model (just outside the 20-day window). The source of this stream is labeled as coronal hole \#2 in the photosphere map at the bottom. The model proton number density and temperature also generally agree with the SWEAP data away from the discrepancies caused by the offset of the two high-speed streams. 

Figure \ref{fig4} shows the radial components of the model magnetic field and velocity, proton density and temperature compared with OMNI data at Earth and PSP data for the second orbit around the Sun from 2019/01/20 to 2019/06/18. The model radial magnetic field compares reasonably to OMNI data throughout the entire period in terms of peak strengths and periodic polarity changes. Apparently, Earth traversed through negative sectors much longer than through positive sectors during this period in contrast to the first PSP orbit when the opposite was observed. This makes sense because Earth was mostly above/below the equatorial plane during PSP's first/second orbit. On the other hand, there are some discrepancies between the model and the observed radial velocities, particularly around 2019.09, 2019.14, 2019.17, 2019.32, 2019.37, and 2019.40. We note that the discrepancies around 2019.37 are not a result of any inaccuracies that may be present in the boundary conditions, but rather due to the passage of CMEs on 2019/05/11 and 2019/05/14, which the model does not account for. The model proton density and temperature also agree reasonably with OMNI data, except for the noted times when the discrepancy between the model and observed radial velocities are significant.

In the first half of the period up to 2019.21, the model radial magnetic field and velocity at PSP fluctuate mostly in the -10 to +10 nT and 400 to 600 km s$^{-1}$ ranges, respectively, as the heliocentric distance gradually decreases to 0.5 au. With the exception of a very fast stream ($>$700 km s$^{-1}$) at 2019.17, these results are consistent with those for the first orbit. During the second solar encounter, the model radial magnetic field increases to +70 nT while velocity fluctuation remains in the 400 to 600 km s$^{-1}$ up to the perihelion before dropping to the 300 to 450 km s$^{-1}$ range until 2019.30. The radial magnetic field changes to -70 nT at DOY 95 as PSP crosses the heliospheric current sheet around the perihelion in the model. After the second solar encounter, the model radial field and velocity fluctuate mostly in the -10 to +10 nT and 300 to 600 km s$^{-1}$ range as PSP gradually approaches the second aphelion. The model proton density and temperature steadily increase from 2-20 cm$^{-3}$ and $5 \times 10^{4}-2 \times 10^{5}$ K near 1 au to 50-500 cm$^{-3}$ and 1-5$\times 10^{5}$ K during the second solar encounter, followed by a steady decrease to aphelion at 0.94 au. These results are mostly consistent with the PSP FIELDS and SWEAP data away from the solar encounter, excluding comparison at distances much larger than 0.25 au as discussed earlier.

As noted above, there are some significant discrepancies between the model and PSP data during the second solar encounter that we must address. The left panel of Figure \ref{fig5} provides an expanded view of the simulation results at PSP for the second solar encounter during the 20-day period around perihelion 2 (2019/04/04 22:40 UT). The model suggests that PSP crosses the heliospheric current sheet from positive to negative magnetic polarity at 2019/04/04 13:12 UT (DOY 94.6) and then remains in the negative sector after the perihelion. On the other hand, the FIELDS data indicate that PSP encountered mostly negative magnetic polarity throughout the entire 20-day period. While PSP observed strictly slow wind streams between 230 and 450 km s$^{-1}$, the model velocity fluctuates between 300 and 650 km s$^{-1}$. The right panel of Figure \ref{fig5} shows a snapshot of 3D and spherical slices (35.7 R$_{s}$ or 0.166 au) of the model magnetic field polarity and radial velocity at the perihelion, where the location of PSP is marked by an `x'. These plots suggest that PSP was still within 2\textdegree\ of the heliospheric current sheet 10 hours after the crossing in the model. The radial velocity plots show that PSP navigated through the middle of a low speed band surrounding the heliospheric current sheet that originated near the boundary of the southern polar coronal hole, as indicated at the bottom. After the perihelion, PSP remained below the heliospheric current sheet in this low-speed band until the end of the solar encounter, which is largely consistent with observations.

We find the largest discrepancies between the model and observations over the 10 days leading to the perihelion, where a high-speed stream (650 km s$^{-1}$) of positive magnetic polarity, which was never observed by PSP, appears in the model. Figure \ref{fig6} shows 3D plots and spherical slices (63.8 R$_{s}$ or 0.297 au) of the model magnetic field polarity and radial velocity at 2019/03/28 20:00:00 UT (DOY 87.8), where the location of PSP is marked by an `x'. These plots suggest that PSP was 15\textdegree\ above the heliospheric current sheet, which disagrees with FIELDS observations of mostly negative magnetic polarity at that time. This high-speed stream of positive polarity at PSP is traced to a southward extension of the northern polar coronal hole as shown at the bottom of the right panel of Figure \ref{fig6}. On the contrary, the steady slow streams of predominantly negative magnetic polarity observed by PSP most likely originated from the edge of the southern polar coronal hole as marked on the bottom plot.

The left panel of Figure \ref{fig7} shows the model prediction for the third solar encounter during the 20-day period around perihelion 3 (2019/09/01 17:50 UT). Since the HMI-ADAPT-WSA maps are only available up to 2019/08/13 20:00 UT at the time of this simulation, we extend the MHD calculations by rotating the last boundary frame at the solar rotation rate, assuming that the solar wind conditions persist over the next 3-4 weeks. The model suggests that PSP remains in the negative sector below the heliospheric current sheet throughout the entire period as the radial field steadily increases in strength from -20 nT at 0.348 au to -76 nT at 0.166 au and then back to -20 nT at the end of the time window. The radial velocity fluctuates mainly between 300 and 400 km s$^{-1}$, except for a high-speed stream above 550 km s$^{-1}$ at DOY 245 just one day after the perihelion. The proton density and temperature also gradually increase from 50-100 cm$^{-3}$ and 1-2$\times 10^{5}$ K at 0.3-0.35 au to 100-300 cm$^{-3}$ and 2-6.5$\times 10^{5}$ K near the perihelion. The right panel of Figure \ref{fig7} shows a snapshot of 3D and spherical slices (35.7 R$_{s}$ or 0.166 au) of the model magnetic field polarity and radial velocity at the perihelion, where the location of PSP is marked by an `x'. These plots suggest that PSP is 10\textdegree\ below the heliospheric current sheet at the closest approach to the Sun when PSP briefly reaches the edge of a low speed band surrounding the heliospheric current sheet that originated near the boundary of the southern polar coronal hole, as indicated on the bottom plot. 

\section{Summary and Discussion}

Using time-varying boundary conditions derived from ADAPT-WSA model with SDO/HMI magnetograms, we performed a 3D time-dependent MHD simulation of the inner heliosphere for the first two PSP orbits. These boundary conditions were chosen to ensure the best (most reasonable) agreement between the model and near-Earth solar wind data at 1 au as discussed in the results section. The MS-FLUKSS model output along the first PSP orbit compare reasonably with FIELDS and SWEAP data where signal/noise ratios are sufficiently high. During the first solar encounter, the model suggests that PSP was magnetically connected to a southern equatorial coronal hole before crossing the heliospheric current sheet from negative to positive sector shortly after the perihelion, which agrees with observations and other models discussed by \cite{Riley2019ApJ,Badman2019ApJS,Szabo2019ApJS}. The model suggests that the solar wind streams sampled by PSP during this time were primarily connected with two equatorial coronal holes of opposite magnetic polarity.

On the other hand, the model deviates from PSP observations during the first half of the second solar encounter, where it presents a high-speed stream above 650 km s$^{-1}$ and of positive magnetic polarity just 7 days prior to the perihelion that was never detected by the spacecraft. Moreover, the model indicates that PSP would cross the heliospheric current sheet near the perihelion from positive to negative sector, whereas the observed magnetic field direction remained radially inward and most likely connected to the southern polar coronal hole throughout the second encounter. To identify the source of this error, we must consider the longitude separation of Earth and PSP as the latter faces the far side of the Sun during the solar encounter. Apparently, there is an active region that emerges between 2019/03/20 and 2019/03/24 that undergoes significant evolution after leaving the magnetograph's field of view.  When it moves back into the field of view around midday 2019/04/09, it significantly alters the streamer belt configuration of the model. This uncertainty suggests that the model most likely contains errors for at least several days prior to 2019/04/09, which may have been responsible for the large discrepancies at PSP leading up to the perihelion on 2019/04/04.

Next, the model predicts predominantly low-speed streams of negative magnetic polarity connected to the southern polar coronal hole throughout PSP's third solar encounter between 2019/08/22 and 2019/09/11. These predictions appear very similar to what the spacecraft observed during the previous solar encounter. We note that the model used boundary conditions from nearly 3 weeks before the third perihelion (2019/09/01) to make the predictions assuming that the solar wind conditions would not change significantly over the next solar rotation. However, the solar wind structure can change unexpectedly sometimes even during the current low-activity period near the solar minimum. Thus, we will update these initial predictions with newer boundary conditions later on. We will also consider several possible improvements to the model before the next prediction runs. For example, we determined the best input magnetograms based on comparison of the WSA model with near-Earth data in the current study as is customary, but the outcome may not be necessarily best for comparing at PSP, particularly during the solar encounters when the spacecraft mostly faced the far side of the Sun. Hence, we look to take PSP data into account to select the best input magnetograms in future studies. The height of the source surface of the PFSS model, which is a free parameter set to 2.5 R$_s$ in this study, could also be adjusted to improve the open flux (and other quantities as a result) at 1 au as suggested by \cite{Arden2014JGR}. Finally, we will look into the evolution of solar wind turbulence along the PSP trajectory by solving the Reynolds-averaged MHD equations with turbulence and interstellar pickup ions taken into account, which are fully implemented in MS-FLUKSS \cite[e.g.,][]{Pogorelov2012AIP,Kryukov2012AIP}, in a follow-up study.


%% Included in this acknowledgments section are examples of the
%% AASTeX hypertext markup commands. Use \url without the optional [HREF]
%% argument when you want to print the url directly in the text. Otherwise,
%% use either \url or \anchor, with the HREF as the first argument and the
%% text to be printed in the second.

\acknowledgments

The PSP FIELDS and SWEAP science teams acknowledge NASA contract NNN06AA01C. This work is partly supported by the PSP mission through the UAH-SAO agreement SV4-84017. This work is also supported by the NSF PRAC award OCI-1144120 and related computer resources from the Blue Waters sustained-petascale computing project. Supercomputer time allocations were also provided on SGI Pleiades by NASA High-End Computing Program awards SMD-16-7570 and SMD-18-2194 and on TACC Stampede2 and SDSC Comet by NSF XSEDE project MCA07S033. TKK acknowledges support from NASA grant 80NSSC19K0008 and AFOSR grant FA9550-19-1-0027. This work utilizes data produced collaboratively between Air Force Research Laboratory (AFRL) and the National Solar Observatory. The ADAPT model development is supported by AFRL. The authors acknowledge use of the SDO/HMI data from the \href{http://jsoc.stanford.edu/}{Joint Science Operations Center} and the \href{https://cohoweb.gsfc.nasa.gov/coho/}{SPDF COHOWeb} database for OMNI data. The PSP data used in this study will be released to the scientific community and the public on November 12, 2019 through \href{https://spdf.gsfc.nasa.gov/}{NASA SPDF}, \href{https://spdf.gsfc.nasa.gov/}{SDAC}, the Science Operation Centers of the four science investigation teams, and the APL Parker Solar Probe Gateway. 




%\bibliography{sample63}{}
%\bibliographystyle{aasjournal}



\end{document}